# Raman scattering evidence of hydrohalite formation on frozen yeast cells

K. A. Okotrub[a] and N. V. Surovtsev[a,b,]*

[a]Institute of Automation and Electrometry, Russian Academy of Sciences, Novosibirsk, 630090, Russia,

[b]Novosibirsk State University, Novosibirsk, 630090, Russia,

* Corresponding author

E-mail address: lab21@iae.nsk.su

Fax: (7)-383-3333863

Telephone: (7)-383-3307978

Postal address: pr. Ak. Koptyuga 1, 630090, Novosibirsk, Russia

**Abstract**

We studied yeast cells in physiological solution during freezing by Raman microspectroscopy technique. The purpose was to find out the origin of a sharp peak near ~3430 cm$^{-1}$ in Raman spectrum of frozen mammalian cells, observed earlier (J. Dong et al, Biophys. J., 99 (2010) 2453), which presumably could be used as an indicator of intracellar ice appearance. We have shown that this line (actually doublet of 3408 and 3425 cm$^{-1}$) corresponds to Raman spectrum of hydrohalite (NaCl·2H$_2$O), which is formed as the result of the eutectic crystallization of the liquid solution around the cells. We also show that the spatial distribution of hydrohalite in the sample significantly depends on the cooling rate. At lower cooling rate (1°C/min), products of eutectic crystallization form layer on the cell surface which thickness varies for different cells and can reach ~1 μm in thickness. At higher cooling rate (20°C/min), the hydrohalite distribution appears more homogeneous, in the sample, and the eutectic crystallization layer around the cells was estimated to be less than ~20 nm. These experimental results are consistent with scenarios predicted by the two-factor hypothesis for freezing induced cell injury. This work demonstrates a potential of Raman microspectroscopy to study peculiarities of the eutectic crystallization around single cells *in vivo* with the high spatial resolution.

**Introduction**

Various biological and medical applications involve cell preservation technology ([30] and references therein). Cryopreservation of living cells is used for long term culture fixation to suppress genetic and phenotypic drifts. Cryobanking is used for domestic livestock, endangered species genetic resources storage and artificial insemination. Notably, multiple damaging factors arise during cooling cells to cryogenic temperatures, such as ice formation (both extra- and intracellular ones), dehydratation of cells and their membranes, phase transitions of membranes, the toxic effects of cryoprotectants, and eutectic crystallization. How these factors affect the survival ability of cells remains unanswered. While most cryopreservation protocols have been developed empirically [9], our understanding of physical processes underneath of the cryoprocesses is still obscure. That is why understanding of the cells cryoinjury mechanisms [7,12,16,17,30] is an important task, attracting attention of many research groups.

Several nondestructive techniques have been adapted to investigation of cells freeze/thaw processes investigations *in vivo*, including cryomicroscopy [1,4,25,27,29], infrared spectroscopy [8,21,28], and differential scanning calorimetry [12,14]. However, since some of these techniques provide indirect information or require ambiguous interpretations (see, for example, ref. 27, where the distinction between intercellular ice formation and darkening effects in cryomicroscopy is discussed), it is of great importance to develop other methods to study the cryopreservation process. Raman scattering spectroscopy is a non-invasive, nondestructive technique, which is sensitive to chemical composition and phase state, and, thus, is exceptionally well suited for these kinds of problems. Raman spectroscopy is actually used with rapidly growing popularity to study various biological problems (e.g. [6,13,20,22,23,26,31] and others). However, until very recently there have been no Raman scattering studies on cryobiology of the cells. To the best of our knowledge, a recent work [10] was the pioneering one in this direction, and this work is an excellent illustration of the Raman technique capabilities for identification of extra- and intracellular ice, spatial distribution of a cryoprotectant and the responses of subcellular structures at the frozen state.

Among others results, authors of the work [10] reported an observation of a sharp peak near ~3430 cm$^{-1}$ in intracellular ice (Fig. 1 B of [10]). Origin of this line was not discussed in the work except for the notion that it could be due to trace amounts of organic matter. However, it is important to find out which substance is responsible for the line at 3430 cm$^{-1}$. Another motivation to pay attention to this line is the assumption that the line indicates the intracellular ice [10]. If true, the line could serve as an excellent indicator of intracellular ice appearance during cell freezing. Here, we describe a detailed study of the freezing process of *Saccharomyces cerevisiae* yeast cells by Raman microspectroscopy. We demonstrate that the line at 3425 cm$^{-1}$ in Raman spectrum is due to hydrohalite (NaCl·2H$_2$O) layer on the cell surface. Measuring Raman spectra, we estimate the thickness of the layer quantitatively, and demonstrate that the thickness depends on the cooling rate.

**Materials and Methods**

*Sample preparation*

Commercially available *Saccharomyces cerevisiae* instant yeast cells pellets (20 mg) were added to 5 ml of isotonic saline solution (0.9 wt % NaCl). For each Raman experiment, one microliter of the cell suspension was placed into a hermetic chamber with silica window (280 μm in thickness).

*Cryostat and temperature control*

Raman scattering experiments were carried out in a home-made cryostat, which is continuously cooled by the flow of liquid nitrogen vapor. The cryostat offers an operating temperature range of 100–320 K with sample in vacuum. The cryostat window is made of silica glass (280 μm in thickness). The cryostat allows us to use microscope objectives with working distance 4 mm and above. Controlling thermocouple was installed close to the sample chamber. The sample was kept at constant temperature when Raman spectra were acquired. The temperature in the laser illuminated area at low temperatures was additionally controlled by the measurement of the position of the ice Raman line (~3100 cm$^{-1}$), which is temperature dependent. An optical closed-cycle helium cryostat (DE-204; Advanced Research Systems, Allenton, PA) was used in a calibration experiment in which temperature dependence of Raman line of ice was measured. The additional temperature control allowed us to take into account possible effects of laser heating of the sample, which did not exceed ~2 °C.

*Confocal Raman microspectroscopy*

Our set-up for Raman experiment with the lateral resolution of ~1 μm includes a confocal microscope, based on a modified Orthoplan microscope (Leitz, Wetzlar, Germany), and an SP2500i monochromator (Princeton Instruments, Acton, MA), equipped with a Spec-10:256E/LN charge-coupled device detector (Princeton Instruments, Trenton, NJ). Wavelength calibration of the spectrometer was done by a neon-discharge lamp. Raman scattering was excited by a solid-state laser (Millennia; Spectra Physics, Santa Clara, CA) at a wavelength of 532.1 nm and 8 mW of power (the power entering to the cryostat). A 100× air objective with NA = 0.75 and working distance 4.6 mm (Leica, Wetzlar, Germany) was used in all measurements. The diameter of a confocal pinhole was 120 μm, which corresponds to a spot of 1.2 μm (diameter) in the focal plane. The optical path from the objective to the sample includes two silica windows (sample chamber and cryostat windows). Longitudinal aberrations, caused by the windows, reduced the longitudinal resolution to ~12 μm (FWHM), while the lateral resolution was about ~1 μm. Point spread function (PSF) was estimated by numerical calculations (PSFlab [19]). Since the size of yeast cells is ~8 μm or less, an extracellular matter contributed to acquired spectra.

*Raman experiment with saline solution*

An aqueous solution with 25 wt % NaCl was used to study the Raman spectrum of ice and hydrohalite after eutectic crystallization. The solution was sealed into a glassy tube. An optical closed-cycle helium cryostat was used for sample cooling/heating. Raman spectra of the opaque white sample (after freezing) were measured in nominally right-angle scattering by a TriVista 777 triple-grating spectrometer (Princeton Instruments, Acton, MA) equipped with a Spec-10:400BR charge-coupled device detector (Princeton Instruments, Trenton, NJ). An Ar-ion laser Stabilite 2017 (Spectra Physics, Santa Clara, CA) at a wavelength of 488 nm and 200 mW of power was used for excitation. Raman scattering was collected from a volume of about 1 mm$^3$.

**Results**

We studied Raman spectrum of the cells during cooling. The most of the experiments was carried out with a cooling rate of 1 ºC/min, which is typical for a number of cryopreservation protocols. Representative Raman spectra of single yeast cells at different temperatures are shown in Fig. 1. The spectrum at room temperature consists of the CH band (2840–3000 cm$^{-1}$) of organic matter and the broad OH band of the liquid water (2900–3600 cm$^{-1}$). Extracellular ice spontaneous nucleation took place in the temperature range –15 – –20 °C (Fig. 1, curve C), which is manifested by the prominent peak centered at 3140 cm$^{-1}$ in Raman spectra. The contribution of the liquid water to Raman spectra vanished at temperatures below –40°C, and new sharp crystalline lines appeared at 3425 and 3545 cm$^{-1}$ (Fig. 1, curve D). Also a weaker line near 3408 cm$^{-1}$ could be distinguished in this curve. The peak near 3425 cm$^{-1}$ is a counterpart of the peak at ~3430 cm$^{-1}$ found in the work [10]. In the Raman spectrum of the extracellular matter (Fig. 2, curve A), only band characteristic for ice was observed.

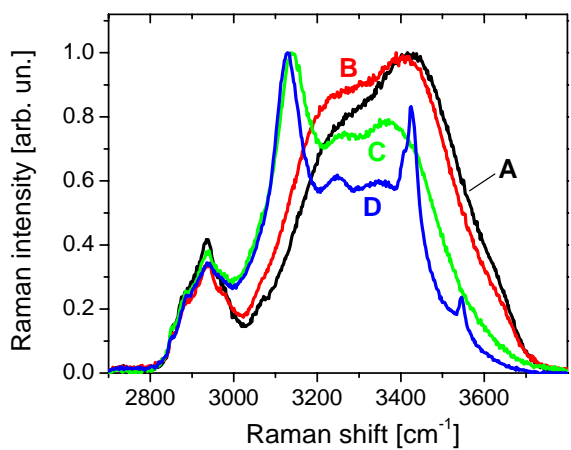 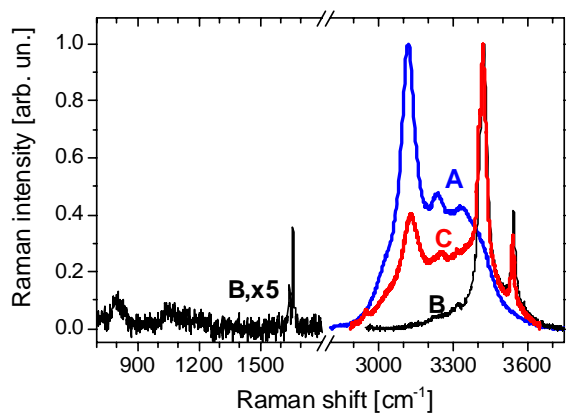

**Fig. 1.** Raman spectra of yeast cells. Curve A – at room temperature (+25 °C); curve B – supercooled water spacemen (–10 °C); curve C – after extracellular ice formation (–20 °C); B – after eutectic crystallization (–46 °C).

**Fig. 2.** Raman spectra of the saline solution after eutectic crystallization. Curve A – extracellular ice at $T = -73$ °C; curve B - hydrohalite at $T = -22$ °C; curve C – macroscopic (~1 mm$^3$) sample of 25 wt % NaCl solution at $T = -50$ °C.

To identify the origin of the observed peaks, we measured a number of aqueous solutions of different organic substance at low temperatures, but did not find peaks like 3425 and 3545 cm$^{-1}$ in Fig. 1. However, these peaks were detected the Raman spectrum of the boundary of a frozen brine droplet. Additionally, we were able to measure Raman spectra of an inclusion in the frozen physiological solution, which had only peaks at 3408, 3425 and 3545 cm$^{-1}$ in the range of 2900 - 3700 cm$^{-1}$ (Fig. 2, curve B). To characterize this inclusion further, we recorded the Raman spectrum in the range 650 - 1800 cm$^{-1}$ (Fig. 2, curve B), where only two sharp lines at 1640 and 1660 cm$^{-1}$ were present. The inclusion melted at temperature about –20 °C. The Raman spectrum of the liquid melt was similar to the NaCl/water solution spectra [11]. Both the Raman spectrum and the melting temperature of the inclusion correspond very well to those of hydrohalite (NaCl·2H$_2$O) [3]. Thus, we conclude that peaks at 3425 and 3545 cm$^{-1}$ in Raman spectrum of frozen yeast cells (Fig. 1) correspond to the appearance of hydrohalite in the illuminated volume of the sample.

Raman spectra were acquired for a number (~30) of yeast cells, frozen at low (1 ºC/min) cooling rate. The hydrohalite contribution to the Raman spectra varied significantly from cell to cell, sometime several fold comparing to the spectrum D in Fig. 1.

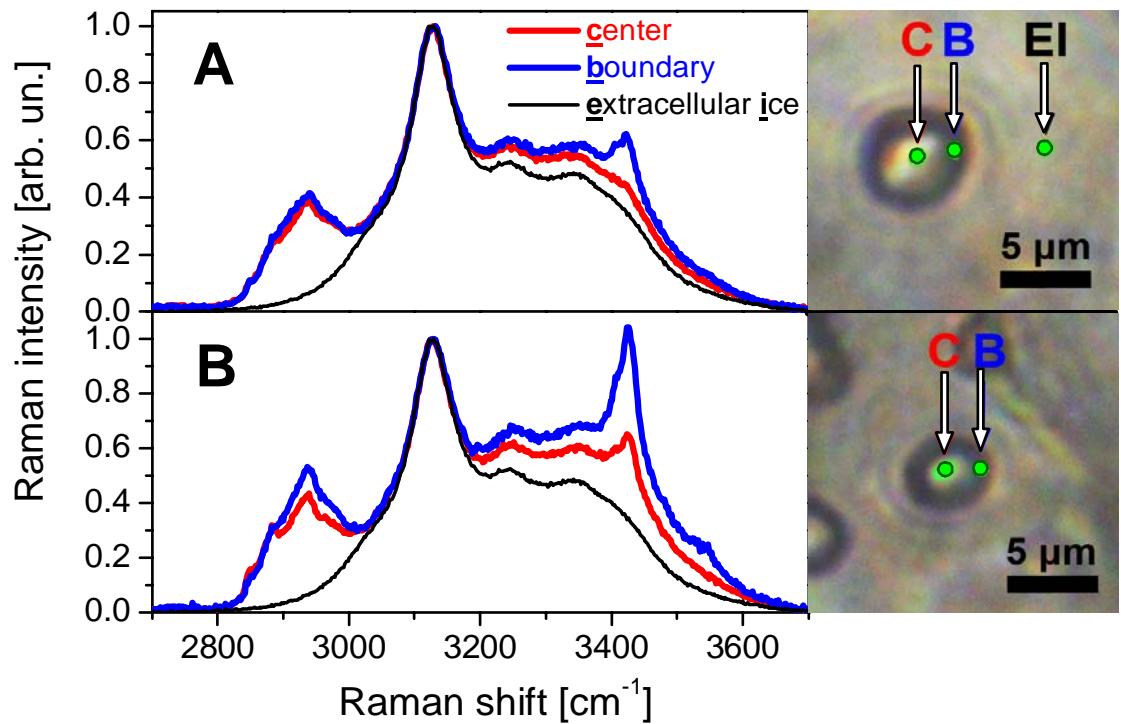

**FIGURE 3.** Raman spectra of the different parts of the yeast cell sample at $T = -48$ °C. The different parts shown in the photo are the center of the frozen cell (labeled by "C"), the cell boundary (labeled by "B"), and the extracellular ice (labeled by "EI"). Top panel is the case of a cell with the low amount of hydrohalite, and bottom panel is the case of a cell with the high amount of hydrohalite.

We also investigated spatial distribution of hydrohalite by recording Raman spectra from the center of yeast cell, from cell boundary, and from extracellular areas (e.g. photo in Fig. 3). Representative Raman spectra from different spatial places are shown in Fig. 3 for a cell with a high portion of hydrohalite (top panel) and for a cell with a low portion of hydrohalite (bottom panel). As it is seen from this figure, hydrohalite lines are more intensive from the boundary than from the cell center. This can be accounted for by hydrohalite formation on surface of frozen yeast cells. The hydrohalite contribution in Raman spectra measured from the cell center can be explained by the hydrohalite from the top and bottom cell boundaries, since illuminated volume, i.e. PSF, exceeds the cell diameter in the longitudinal direction. No significant variations of hydrohalite line intensity were found for different points on the boundary of a particular cell. These results suggest that the hydrohalite forms a rather uniform layer around the cell. Thus, yeast cells frozen at low (1 °C/min) cooling rate are surrounded by a hydrohalite layer.

We next explored the role of cooling rate on hydrohalite formation. To this end, a high cooling rate (~20 °C/min) was applied to the sample. Yeast cells frozen at the low and high cooling rates look differently. In case of the low cooling rate, the cells experienced a significant shrinking during freezing, while at the high cooling rate the cell size remained the same. Visual examination of the samples revealed that the extracellular ice was nearly homogeneous for the low cooling rate, while it was heterogeneous for the high cooling rate. The hydrohalite contribution to the Raman spectra was vanishingly small for the cells frozen at high cooling rate.

Quantification of the hydrohalite to ice volume ratio in the illuminated area of a sample requires an independent calibration. For calibration, the Raman experiment with a sample of 25 wt % NaCl aqueous solution was carried out. To confirm homogeneity of the sample, Raman spectra from different parts of the sample were measured. The Raman line of hydrohalite was obtained from the spectrum after the subtraction of the contribution of the Raman band of ice. As result, the ratio of hydrohalite Raman peak at 3425 cm$^{-1}$ and ice Raman peak at 3130 cm$^{-1}$ was found to be near 2.06 in the temperature range $-30 - -50$ °C (Fig. 2, curve C)

**Discussion**

Our experimental results demonstrate that spectral lines at 3408, 3425, and 3545 cm$^{-1}$ appeared in the Raman spectrum of yeast cells frozen in physiological salt solution correspond to the hydrohalite formation. Since it is usual practice to use a salt buffer solution for cell cryopreservation, the results of our study are important for the cases when buffer solutions with the significant concentration of NaCl are used. It appears that Raman spectroscopy can serve as a non-invasive technique for the control of the hydrohalite appearance on frozen cells.

Concentration of NaCl in physiological solution is 0.9 wt %, but the appearance of extracellular ice at low temperature leads to an increase of NaCl concentration in the liquid part of solution. Eutectic crystallization occurs in the brine with 23.3 wt % NaCl concentration at the temperature -21.2°C [5]. In our experiments, the hydrohalite formation occurred at –40°C, when the solution was highly supercooled. Observed supercooling effect agrees well with the results obtained by differential scanning calorimetry [12]. It was proposed [12] that the eutectic hydrohalite formation affects its viability by two injury mechanism: mechanical damage to membrane due to extracellular eutectic crystallization and intracellular eutectic formation. Our Raman experiments visualize the eutectic hydrohalite formation on yeast cells at low cooling rate.

At the high cooling rate the supercooling effect increases, which decreases the size of ice formation critical cluster. Therefore at the high cooling rate the density of ice crystallites increases. As the result, the ice crystals capture the most of the brine droplets, and the amount of the unfrozen fraction surrounding the cells is reduced. In this case much less hydrohalite is formed on the cells.

The cooling rate of maximal viability for *Saccharomyces cerevisiae* yeast cells was reported being ~7 °C/min [5,18]. Therefore, high (20 °C/min) and low (1 °C/min) cooling rates of the Raman experiment correspond to the domination of the intracellular ice injury and of the hypertonic solution injury, respectively, according to the two-factor hypothesis [15]. The hydrohalite presence on yeast cells indicates exposure of the cell in concentrated saline solution, and vice versa. Therefore, high hydrohalite contribution for the high cooling rate and a vanishing hydrohalite contribution for the low cooling rate, as it is observed, agree well with the two-factor hypothesis.

Raman data presented here can be used for an estimate of the hydrohalite amount. Raman line intensity is proportional to the material volume fraction in the illuminated volume. Thus, the relative volumes of hydrohalite and ice are related to the intensities of their Raman lines by

$$\upsilon_{hh}/\upsilon_{ice} = \xi \cdot (I_{hh}/I_{ice}), \qquad (1)$$

where $\upsilon_{hh}$, $\upsilon_{ice}$ are the volumes, occupied by hydrohalite and ice, within the illuminated volume; $I_{ice}$ is intensity of the Raman OH stretch line of ice (~3125 cm$^{-1}$); $I_{hh}$ is intensity of the Raman line of hydrohalite at 3425 cm$^{-1}$; $\xi$ is a proportionality coefficient, which should be found from a normalization experiment.

The spectrum of 25 wt % NaCl frozen solution shown in curve C in Fig. 2 was used for $\xi$ calibration. This solution provides 27.8 vol % of hydrohalite in the sample. The Raman line of hydrohalite was found from the spectrum (Fig. 2) after the subtraction of ice contribution. Assuming that hydrohalite and ice have a nearly uniform distribution in the illuminating volume, this gives us an estimate of $\xi \approx 0.187$ for the temperature range -20 °C ÷ -50 °C.

Raman spectrum from the center of yeast cells and Eq.(1) can be used to estimate the thickness of the hydrohalite layer, since the cell size is smaller than longitudinal length of the sample volume, from which Raman scattering is measured (FWHM of PSF). Using value $\xi \approx$ 0.187 from the calibration procedure described above and Eq.(1), we can estimate thickness for each individual cell measured. For example, for the cell with the Raman spectra shown as a curve B in bottom panel of Fig. 3 the ratio of hydrohalite and ice volumes estimated as ~ 0.04. Subtracting the diameter of the cell (~3 μm, bottom panel of Fig. 3) from the longitudinal length of the illuminated volume (~12 μm, longitudinal FWHM of PSF), we found the hydrohalite amount would correspond to the layer thickness of ~0.4 μm. Eutectic crystalline structure is composed both of hydrohalite and ice with relative hydrohalite volume of 25.6 %, and there presents two layers above and below the cell. Then, the estimate of eutectic crystalline layer around the cell is ~1 μm in the case of the cell shown in bottom panel of Fig. 3. For the cell shown in top panel of Fig. 3 the estimate of the eutectic crystalline layer gives approximately ten times smaller value. The different contributions of hydrohalite for different cells reflect the random character of the brine volume concentrated around a cell during freezing.

We also made a more precise estimate of the eutectic crystalline layer thickness, taking into account a realistic distribution of laser intensity. However, the estimated value of the eutectic crystalline layer was found to be essentially the same as above.

Interestingly, the thickness of the eutectic crystalline layer differed a lot in the case of the high (20 °C/min) cooling rate. Experimentally, the hydrohalite contribution was very low in this case. Layer thickness estimates give at least 60-fold smaller value comparing to the cell in bottom panel of Fig. 3. Thus, for the high cooling rate the eutectic crystalline layer was estimated to be less than ~20 nm.

**Conclusion**

We investigated physical processes happening during cryopreservation of cells by Raman microspectrscopy using budding yeasts as a model organism. Raman scattering from individual yeast cells in the physiological solution at low temperatures was measured. We show that lines 3408, 3425, and 3545 cm$^{-1}$ appearing in Raman spectrum of frozen yeast cells corresponds to the spectrum of hydrohalite (NaCl ·2H$_2$O). Hydrohalite formation on the cell surface is a result of the eutectic crystallization of the liquid solution surrounding the cells. Based on our experimental data we estimated the thickness of the eutectic crystalline layer. Particular space distribution of eutectic crystallization products depends on cooling rate, and eutectic crystalline layer thickness varies significantly from cell to cell. It is known that the effects of eutectic crystallization (e.g. [2,7,12,24]) are crucially important for the cell viability during freezing. Raman spectroscopy provides the excellent possibility to study the eutectic crystallization, which can be done *in vivo* with single cells.